\documentclass[10pt]{article}
\usepackage{multicol}
\usepackage{graphicx}
\usepackage{amsmath}

\begin{document}

\begin{center}
{\Large \bf Study of kinetic freeze-out parameters as function of rapidity in pp collisions at CERN SPS energies}

\vskip1.0cm

Muhammad~Waqas$^{1,}${\footnote{E-mail: waqas\_phy313@yahoo.com}},
H. M. Chen$^{1,}$, 
Guang Xiong Peng$^{1,}$\footnote{Correspondence: gxpeng@ucas.ac.cn},\\
Abd Al Karim Haj Ismail$^{2,3}$\footnote{Correspondence: a.hajismail@ajman.ac.ae},
Muhammad Ajaz$^{4,}$, 
Zafar wazir$^{5}$, 
Ramoona Shehzadi $^{6}$, 
Sabiha Jamal $^{4}$, 
Atef AbdelKader$^{2,3}$

\vskip.25cm

{\small\it $^1$School of Nuclear Science and Technology, University of Chinese Academy of Sciences, Beijing 100049, Peoples Republic of China,
\\
$^2$Department of Mathematics and Science, Ajman University, PO Box 346, UAE,
\\
$^3$Nonlinear Dynamics Research Center (NDRC), Ajman University, PO Box 346, UAE,
\\
$^4$Department of Physics, Abdul Wali Khan University Mardan, 23200 Mardan, Pakistan,
\\
$^5$Department of physics, Ghazi University, Dera Ghazi Khan, Pakistan,
\\
$^6$Department of Physics, University of the Punjab, Lahore, Pakistan}
\end{center}

\vskip1.0cm

{\bf Abstract:} We used the blast wave model with Boltzmann Gibbs
statistics and analyzed the experimental data measured by NA61/SHINE Collaboration
in inelastic (INEL) proton-proton collisions at different rapidity slices at different
center-of-mass energies. The particles used in this study are $\pi^+$, $\pi^-$, $K^+$,
$K^-$ and $\bar p$. We extracted kinetic freeze-out temperature, transverse flow velocity
and kinetic freeze-out volume from the transverse momentum spectra of the particles. We observed
that the kinetic freeze-out temperature is rapidity and energy dependent, while transverse flow velocity
does not depend on them. Furthermore, we observed that the kinetic freeze-out volume is energy dependent
but it remains constant with changing the rapidity. We also observed that all these three parameters
are mass dependent. In addition, with the increase of mass, the kinetic freeze-out temperature increases,
and the transverse flow velocity as well as kinetic freeze-out volume decreases.
\\



\vskip1.0cm
\begin{multicols}{2}

{\section{Introduction}}

To study the dynamics of high energy collisions, the transverse momentum spectra of the particles
are the most important tool to be used. The proton-proton (pp) interactions are used as the baseline
and are significant for understanding the particle production mechanism [1].

In order to understand the nuclear reaction mechanism and evolution characteristic, the excitation
functions of some physical quantities (i.e different temperatures including initial temperature ($T_i$), effective
temperature ($T$), chemical freeze-out temperature ($T_{ch}$) and kinetic freeze-out temperature ($T_0$) and mean
transverse momentum $<p_T>$) are very important. We can learn more about the high energy collision process
by analyzing the excitation functions of $T_0$, $<p_T>$, $T_i$, transverse flow velocity ($\beta_T$) and kinetic freeze-out
volume ($V$), in which the excitation functions of the above parameters can be extracted from the transverse
momentum spectra ($p_T$) of the particles. The parameters mentioned above are discussed in detail in Ref. [2--7].
However, we would like to further discuss the temperature here, because temperature is one of the most significant
concepts in thermodynamics and statistical mechanics [8], and it describes the excitation degree of the interacting
system in sub-atomic physics.

There are different kinds of temperature discussed in the literature [3--7], which corresponds to different collision
stages and it is expected that they vary at different stages due to the system evolution. In the present work, we
will study the final state particles, therefore we are interested in kinetic freeze-out temperature which occurs at the stage
of kinetic freeze-out. In literature, there are various kinetic freeze-out scenarios including single, double, triple and
multiple kinetic freeze-out scenarios [9--14]. In addition, there are various claims and dependencies of kinetic freeze-out
temperature on centrality [5--10, 13, 15--17] and collision energy [2, 3, 10, 18]. In our recent work, it is observed that
the kinetic freeze-out temperature also depends on coalescence and isospin symmetry [19]. We wonder if $T_0$ is also
expected to depend on rapidity. For this purpose, we analyze the final state identified particles in different rapidity slices
at different energies in pp collisions and extracted the kinetic freeze-out temperature, transverse flow velocity
($\beta_T$)and kinetic freeze-out volume from the transverse momentum spectra of the particles by using blast wave
model with Boltzmann Gibbs statistics [20, 21].

The structure of $\beta_T$ and $V$ is very complex and they are studied in various publications. For centrality
dependence, most of the studies agree that $\beta_T$ and $V$ decrease with decreasing centrality [4,6,7,13,18], but
the dependence of $\beta_T$ on collision energy is a matter of contradiction [10, 17, 18].

The remainder of the paper includes formalism and method, results and discussions, and conclusions.
\\

{\section{The model and method}}

It is experimentally established that the system which is produced during high energy collision
has an azimuthal anisotropy due to the difference in flow velocities along various directions.
This azimuthal anisotropy occurs due to some initial state geometric effects that rise during
the collision process. Therefore, it can be testified that the departing particles must carry some
impressions of such effects that can have an impact on the nature of the spectra. In order to
include such effects, various models are suggested [22--32]. In the present work we used the blast
wave model with Boltzmann Gibbs statstics (BGBW) [20,21]. Blast wave model is a hydrodynamical based model.
It includes the random thermal motion as well as the flow properties of the particles. Including the
azimuthal anisotropy, the blast wave model gives a complete picture of the quark gluon plasma (QGP)
evolution dynamics. The transverse momentum spectrum of BGBW is given as
\begin{align}
f(p_T)=&\frac{1}{N}\frac{dN}{dp_T} = \frac{1}{N}\frac{gV}{(2\pi)^2} p_T m_T \int_0^R rdr \nonumber\\
& \times I_0 \bigg[\frac{p_T \sinh(\rho)}{T} \bigg] K_1
\bigg[\frac{m_T \cosh(\rho)}{T} \bigg],
\end{align}
where $N$ represents the number of particles, $g$ is the spin degeneracy factor of the particle which is different for different particles (based on $g_n$=2$S_n$+1,  while $S_n$ is the spin of the particle), $V$ denotes the volume of the system under consideration, $m_T$ ($m_T=\sqrt{p_T^2+m_0^2}$) is the transverse mass, $I_0$ and $K_1$ are the modified Bessel functions, $\rho= \tanh^{-1} [\beta(r)]$, $\beta(r)= \beta_S(r/R)^{n_0}$ is the transverse radial flow of the thermal source at radius $0 \leq \ r \leq \ $$R$ with surface velocity $\beta_S$. We used $n_0$ = 2 to be compatible with ref. [20] in the present work, which it closely resembles hydrodynamic profile as mentioned in ref. [20], and results in $\beta_T$ = 0.5$\beta(S)$. Because the maximum value of $\beta(S)$ is 1c, the maximum value of $\beta_T$ is 0.5c. If $n_0$ = 1, it results in $\beta_T$ = (2/3)$\beta(S)$. Therefore, the maximum $\beta_T$ is (2/3)c. If $n_0$ is used to be a non-integer from that less than 1 to above 2, then it corresponds to the centrality from center to periphery. This can lead to a large fluctuation in $\beta_T$. The value of $n_0$ =1 or 2 does not effect the result because it is not very sensitive, but if it is taken as a free parameter, then it leads to a large fluctuation in $\beta_T$, and naturally a small change in $T_0$ occurs. In general, $\beta_T=(2/R^2)\int_0^R r\beta(r)dr=2\beta_S/(n_0+2)=2\beta_S/3$.

We can use Eq. (1) for the fitting of $p_T$ spectra and extract the parameters $T_0$, $\beta_T$ and $V$. Eq.(1) is applicable in only soft $p_T$ ($p_T$=2-3 GeV/c) regime
of the $p_T$ spectra and is valid for a narrow $p_T$ range. In the range of $p_T$$>$3 GeV/c, a hard scattering process should be brought into consideration, which is studied in our previous studies [5, 8, 9, 13, 16, 18].
\\

{\section{Results and discussion}}

The transverse momentum ($p_T$) spectra of $\pi^+$ and $\pi^-$ produced in inelastic (INEL) proton-proton (pp)
collisions [33] at different rapidity slices at different energies are represented in fig. 1 and fig. 2, respectively.
The symbols represent the experimental data of NA61/SHINE Collaboration at CERN (European Council
for Nuclear Research) and different symbols represents the $p_T$ spectra of the particles at different rapidity slices.
The collision energy and rapidity slices are labeled in each panel. The solid curves on the experimental data are our fitting by using
Eq. (1). Different panels corresponds to different collision energies. The lower layer in each panel represents the corresponding ratio
of data/fit. The related values of free parameters, and $\chi^2$ and degrees of freedom (dof) are presented in table 1.
One can see that Eq. (1) provides an approximately well fitting of the data at all rapidity slices.

Fig 3. and fig. 4 are similar to fig. 1 and 2, but they show the transverse momentum spectra of $K^+$ and
$K^-$ respectively at different rapidity slices in pp collisions at different energies. The symbols represent the
experimental data of NA61/SHINE Collaboration [33] measured at CERN. The collision energy and rapidity slices are presented in
\begin{figure*}[htbp]
\begin{center}
\includegraphics[width=16.cm]{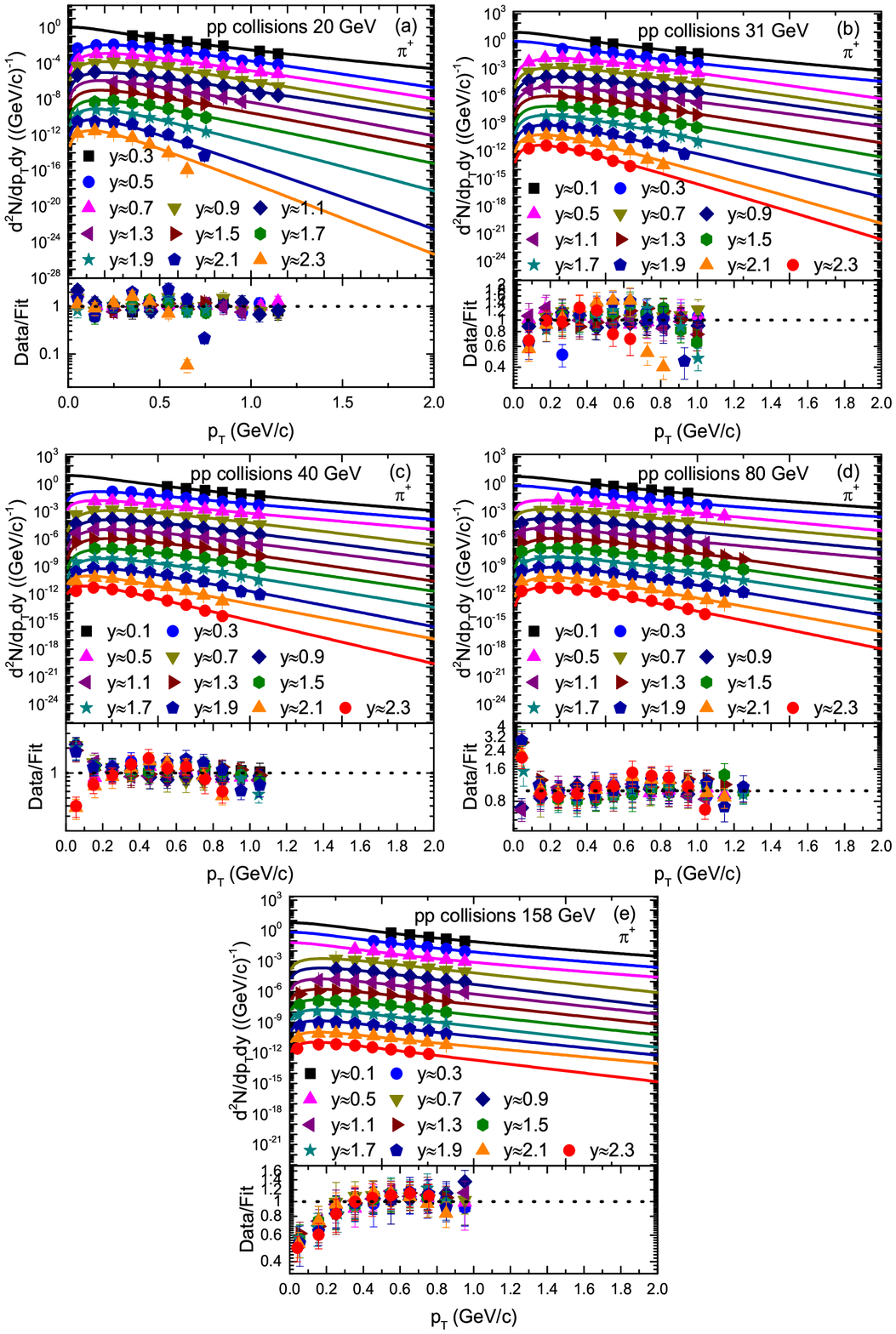}
\end{center}
Fig. 1. Transverse momentum spectra of $\pi^+$ produced in different rapidity
slices in pp collisions. Panel (a)-(e) corresponds to 20, 31, 40, 80 and 158 GeV energy respectively.
The symbols represent the experimental data of NA61/SHINE Collaboration measured at CERN [33]. The
curves are the results of our fits by the Blast Wave model with Boltzmann Gibbs
statistics. The corresponding data/fit ratios are are followed in each panel.
\end{figure*}

\begin{table*}
{\scriptsize Table 1. Values of free parameters ($T_0$ and
$\beta_T$), normalization constant ($N_0$), $\chi^2$, and degree
of freedom (dof) corresponding to the curves in Figs. 1--6.
\vspace{-.50cm}
\begin{center}
\begin{tabular}{cccccccccc}\\ \hline\hline
Collisions       & Rapidity       & Particle     & $T_0$ (GeV)     & $\beta_T$ (c)    & $V (fm^3)$     & $N_0$      & $\chi^2$/ dof \\ \hline
Fig. 1           & y=0.3          & $\pi^+$      &$0.076\pm0.005$  & $0.410\pm0.010$  & $1000\pm100$   & $9.5E-7\pm5E-8$     & 5/4\\
  p-p            & y=0.5          & --           &$0.070\pm0.004$  & $0.280\pm0.007$  & $1000\pm108$   &$5.0E-7\pm4E-8$      & 4/8\\
 20 GeV          & y=0.7          &--           &$0.066\pm0.005$  & $0.284\pm0.010$  & $900\pm90$     & $6E-8\pm6E-9$       & 3.5/8\\
                 & y=0.9          & --         &$0.061\pm0.006$  & $0.284\pm0.011$  & $920\pm120$    & $7E-9\pm6E-10$      & 9/8\\
                 & y=1.1          & --        &$0.056\pm0.004$  & $0.328\pm0.008$  & $900\pm105$    & $4.5E-10\pm4E-11$   & 11/7\\
                 & y=1.3          & --       &$0.051\pm0.004$  & $0.312\pm0.012$  & $800\pm90$     & $6.4E-11\pm5E-12$   & 4/4\\
                 & y=1.5          & --       &$0.048\pm0.006$  & $0.320\pm0.010$  & $860\pm85$     & $4.9E-12\pm5E-13$   & 1/2\\
                 & y=1.7          & --         &$0.045\pm0.004$  & $0.313\pm0.007$  & $870\pm80$     & $3.8E-13\pm4E-14$   & 9/3\\
                 & y=1.9          & --         &$0.041\pm0.005$  & $0.270\pm0.011$  & $800\pm70$     & $4E-14\pm6E-15$     & 5/4\\
                 & y=2.1          & --         &$0.038\pm0.004$  & $0.200\pm0.013$  & $810\pm200$    & $2.5E-15\pm4E-16$   & 118/4\\
                 & y=2.3          & --         &$0.034\pm0.005$  & $0.200\pm0.008$  & $800\pm150$    & $1.5E-16\pm7E-17$   & 120/3\\
 \cline{2-8}
Fig. 2           & y=0.3          &        $\pi^-$  &$0.079\pm0.006$  & $0.421\pm0.013$  & $1000\pm100$   & $6.5E-7\pm6E-8$     & 3/3\\
  p-p            & y=0.5          & --       &$0.070\pm0.005$  & $0.280\pm0.008$  & $1000\pm98$    & $3.5E-7\pm5E-8$      & 6/5\\
   20 GeV          & y=0.7          &--        &$0.067\pm0.006$  & $0.282\pm0.011$  & $920\pm70$     & $4E-8\pm4E-9$       & 3/6\\
                 & y=0.9          & --       &$0.061\pm0.004$  & $0.278\pm0.009$  & $920\pm110$    & $4.5E-9\pm6E-10$    & 7/4\\
                 & y=1.1          & --       &$0.059\pm0.005$  & $0.334\pm0.011$  & $900\pm115$    & $7.2E-11\pm7E-12$   & 8/2\\
                 & y=1.3          & --       &$0.053\pm0.006$  & $0.323\pm0.010$  & $830\pm69$     & $3.6E-11\pm4E-12$   & 3/0\\
                 & y=1.5          & --       &$0.050\pm0.005$  & $0.319\pm0.012$  & $854\pm93$     & $2.2E-12\pm6E-13$   & 0.2/-\\
                 & y=1.7          & --       &$0.047\pm0.006$  & $0.319\pm0.009$  & $857\pm75$     & $1.5E-13\pm5E-14$   & 1/-\\
                 & y=1.9          & --       &$0.043\pm0.005$  & $0.276\pm0.010$  & $808\pm75$     & $1.5E-14\pm5E-15$   & 1/-\\
                 & y=2.1          & --      &$0.040\pm0.006$  & $0.210\pm0.011$  & $819\pm80$     & $1E-15\pm5E-16$     & 10/-\\
                 & y=2.3          & --       &$0.034\pm0.004$  & $0.200\pm0.008$  & $800\pm110$    & $6E-17\pm7E-18$     & 1/-\\
  \cline{2-8}
  Fig. 1         & y=0.1          & $\pi^+$      &$0.088\pm0.006$  & $0.410\pm0.011$  & $1400\pm114$   & $7E-6\pm4E-7$       & 0.1/1\\
  p-p            & y=0.3          & --          &$0.083\pm0.005$  & $0.410\pm0.009$  & $1200\pm210$   & $7.9E-7\pm5E-8$     & 21/4\\
  31 GeV         & y=0.5          &--          &$0.078\pm0.004$  & $0.256\pm0.008$  & $1400\pm100$   & $4.2E-7\pm3E-8$     & 1.5/7\\
                 & y=0.7          &--          &$0.075\pm0.005$  & $0.268\pm0.012$  & $1200\pm108$   & $4.8E-8\pm4E-9$     & 2/7\\
                 & y=0.9          & --         &$0.071\pm0.004$  & $0.293\pm0.009$  & $1300\pm90$    & $4.8E-9\pm4E-10$    & 2/7\\
                 & y=1.1          & --        &$0.067\pm0.006$  & $0.303\pm0.011$  & $1230\pm100$   & $4.2E-10\pm5E-11$   & 2/6\\
                 & y=1.3          & --        &$0.063\pm0.005$  & $0.297\pm0.010$  & $1020\pm110$   & $4.5E-11\pm4E-12$   & 3.5/5\\
                 & y=1.5          & --       &$0.060\pm0.005$  & $0.279\pm0.008$  & $1140\pm105$   & $3.1E-12\pm4E-13$   & 8/5\\
                 & y=1.7          & --       &$0.056\pm0.005$  & $0.260\pm0.009$  & $1000\pm70$    & $3.6E-13\pm4E-14$   & 31/6\\
                 & y=1.9          & --       &$0.052\pm0.004$  & $0.233\pm0.012$  & $1070\pm83$    & $2.7E-14\pm4E-15$   & 25/6\\
                 & y=2.1          & --       &$0.048\pm0.004$  & $0.190\pm0.010$  & $978\pm76$     & $2.6E-15\pm5E-16$   & 80/5\\
                 & y=2.3          & --        &$0.044\pm0.005$  & $0.204\pm0.007$  & $1000\pm104$   & $1.8E-16\pm4E-17$   & 10/3\\
\cline{2-8}
Fig. 2           & y=0.1          & $\pi^-$   &$0.090\pm0.005$  & $0.425\pm0.010$  & $1420\pm104$   & $5.5E-6\pm6E-7$     & 0.2/-\\
  p-p            & y=0.3          & --       &$0.086\pm0.006$  & $0.436\pm0.011$  & $1250\pm110$   & $5.9E-7\pm4E-8$     & 16/-\\
  31 GeV         & y=0.5          &--        &$0.082\pm0.006$  & $0.253\pm0.012$  & $1423\pm109$   & $3.4E-7\pm5E-8$     & 8/5\\
                 & y=0.7          &--        &$0.078\pm0.006$  & $0.263\pm0.008$  & $1216\pm128$   & $3.8E-8\pm7E-9$     & 11/5\\
                 & y=0.9          & --       &$0.073\pm0.006$  & $0.286\pm0.012$  & $1310\pm80$    & $3E-9\pm5E-10$      & 10/4\\
                 & y=1.1          & --       &$0.069\pm0.005$  & $0.324\pm0.012$  & $1255\pm90$    & $7E-11\pm7E-12$     & 2/3\\
                 & y=1.3          & --       &$0.067\pm0.006$  & $0.288\pm0.009$  & $1057\pm100$   & $2.5E-11\pm5E-12$   & 0.3/1\\
                 & y=1.5          & --       &$0.063\pm0.006$  & $0.286\pm0.009$  & $1166\pm112$   & $2E-12\pm6E-13$     & 3.5/1\\
                 & y=1.7          & --       &$0.058\pm0.006$  & $0.270\pm0.012$  & $1040\pm80$    & $2E-13\pm4E-14$     & 6/1\\
                 & y=1.9          & --       &$0.057\pm0.006$  & $0.229\pm0.009$  & $1070\pm92$    & $1.4E-14\pm4E-15$   & 6/1\\
                 & y=2.1          & --       &$0.050\pm0.005$  & $0.190\pm0.008$  & $1000\pm93$    & $1.2E-15\pm4E-16$   & 10/1\\
                 & y=2.3          & --       &$0.044\pm0.005$  & $0.199\pm0.007$  & $1040\pm104$   & $9.4E-17\pm4E-18$   & 9/-\\
\cline{2-8}
Fig. 1           & y=0.1          & $\pi^+$    &$0.097\pm0.005$  & $0.415\pm0.010$  & $1700\pm100$   & $6.7E-6\pm5E-7$     & 0.1/2\\
  p-p            & y=0.3          & --         &$0.093\pm0.005$  & $0.385\pm0.007$  & $1700\pm103$   & $1E-7\pm4E-8$       & 3/5\\
  40 GeV         & y=0.5          &--          &$0.090\pm0.005$  & $0.385\pm0.009$  & $1600\pm106$   & $1.1E-7\pm4E-8$     & 3.5/6\\
                 & y=0.7          &--          &$0.086\pm0.006$  & $0.275\pm0.011$  & $1575\pm100$   & $3.8E-8\pm5E-9$     & 14/7\\
                 & y=0.9          & --         &$0.082\pm0.006$  & $0.276\pm0.008$  & $1540\pm110$   & $3.7E-9\pm4E-10$    & 10/7\\
                 & y=1.1          & --         &$0.078\pm0.004$  & $0.292\pm0.010$  & $1633\pm100$   & $3.4E-10\pm4E-11$   & 6/7\\
                 & y=1.3          & --         &$0.074\pm0.005$  & $0.269\pm0.012$  & $1400\pm90$    & $3.5E-11\pm4E-12$   & 4/7\\
                 & y=1.5          & --        &$0.070\pm0.004$  & $0.290\pm0.010$  & $1437\pm96$    & $2.6E-12\pm4E-13$   & 3/6\\
                 & y=1.7          & --        &$0.067\pm0.004$  & $0.265\pm0.007$  & $1500\pm67$    & $2.1E-13\pm5E-14$   & 21/7\\
                 & y=1.9          & --        &$0.063\pm0.006$  & $0.230\pm0.008$  & $1394\pm100$   & $1.8E-14\pm4E-15$   & 20/7\\
                 & y=2.1          & --        &$0.059\pm0.005$  & $0.285\pm0.012$  & $1400\pm70$    & $4.5E-15\pm5E-16$   & 58/5\\
                 & y=2.3          & --         &$0.055\pm0.006$  & $0.235\pm0.008$  & $1400\pm80$    & $3.2E-16\pm5E-17$   & 41/5\\
  \cline{2-8}
  Fig. 2           & y=0.1          & $\pi^-$   &$0.098\pm0.006$  & $0.418\pm0.010$  & $1700\pm100$   & $4.6E-6\pm4E-7$     & 0.6/-\\
  p-p            & y=0.3          & --        &$0.095\pm0.006$  & $0.385\pm0.009$  & $1720\pm105$   & $8E-7\pm6E-8$       & 1/3\\
  40 GeV         & y=0.5          &--         &$0.091\pm0.005$  & $0.384\pm0.009$  & $1611\pm100$   & $7.6E-8\pm7E-9$     & 2/4\\
                 & y=0.7          &--         &$0.087\pm0.004$  & $0.269\pm0.012$  & $1584\pm110$   & $2.5E-8\pm6E-9$     & 7/5\\
                 & y=0.9          & --        &$0.084\pm0.005$  & $0.270\pm0.008$  & $1500\pm100$   & $2.6E-9\pm5E-10$    & 7/5\\
                 & y=1.1          & --        &$0.080\pm0.005$  & $0.285\pm0.009$  & $1603\pm90$    & $2.1E-10\pm6E-11$   & 14/4\\
                 & y=1.3          & --        &$0.076\pm0.006$  & $0.259\pm0.010$  & $1380\pm85$    & $2.4E-11\pm5E-12$   & 10/3\\
\hline
\end{tabular}%
\end{center}}
\end{table*}

\begin{table*}
{\scriptsize Table 1. Continue.
\vspace{-.50cm}
\begin{center}
\begin{tabular}{cccccccccc}\\ \hline\hline
Collisions       & Rapidity       & Particle  & $T_0$ (GeV)     & $\beta_T$ (c)    & $V (fm^3)$     & $N_0$      & $\chi^2$/ dof \\ \hline
				& y=1.5          & --        &$0.072\pm0.005$  & $0.287\pm0.011$  & $1415\pm73$    & $1.6E-12\pm6E-13$   & 4/2\\
                 & y=1.7          & --        &$0.069\pm0.006$  & $0.265\pm0.009$  & $1490\pm71$    & $1.6E-13\pm5E-14$   & 17/2\\
                 & y=1.9          & --        &$0.064\pm0.005$  & $0.228\pm0.010$  & $1381\pm87$    & $1.3E-14\pm7E-15$   & 14/2\\
                 & y=2.1          & --        &$0.061\pm0.004$  & $0.292\pm0.008$  & $1380\pm76$    & $2E-16\pm5E-17$     & 12/1\\
                 & y=2.3          & --        &$0.057\pm0.005$  & $0.235\pm0.013$  & $1370\pm78$    & $1.3E-17\pm5E-18$   & 73/-\\
\cline{2-8}
  Fig. 1         & y=0.1          & $\pi^+$   &$0.108\pm0.006$  & $0.425\pm0.011$  & $2000\pm100$   & $4.9E-6\pm4E-7$     & 0.2/2\\
  p-p            & y=0.3          & --        &$0.104\pm0.004$  & $0.435\pm0.008$  & $1824\pm70$    & $5.2E-7\pm5E-8$     & 1/4\\
  80 GeV         & y=0.5          &--         &$0.100\pm0.005$  & $0.342\pm0.010$  & $1800\pm66$    & $1.1E-7\pm3E-8$     & 1/6\\
                 & y=0.7          &--         &$0.095\pm0.004$  & $0.363\pm0.010$  & $1829\pm100$   & $1.1E-8\pm7E-9$     & 0.4/5\\
                 & y=0.9          & --        &$0.091\pm0.005$  & $0.364\pm0.009$  & $1892\pm110$   & $9.5E-10\pm6E-11$   & 5/6\\
                 & y=1.1          & --        &$0.087\pm0.006$  & $0.379\pm0.012$  & $1850\pm103$   & $8.8E-11\pm7E-12$   & 7/7\\
                 & y=1.3          & --        &$0.082\pm0.005$  & $0.239\pm0.009$  & $1790\pm108$   & $3E-11\pm6E-12$     & 14/9\\
                 & y=1.5          & --        &$0.078\pm0.006$  & $0.259\pm0.008$  & $1700\pm160$   & $3E-12\pm7E-13$     & 11/9\\
                 & y=1.7          & --        &$0.074\pm0.005$  & $0.258\pm0.009$  & $1766\pm70$    & $2.7E-13\pm6E-14$   & 6/9\\
                 & y=1.9          & --        &$0.070\pm0.005$  & $0.255\pm0.011$ & $1800\pm103$    & $2E-14\pm5E-15$     & 15/9\\
                 & y=2.1          & --        &$0.067\pm0.004$  & $0.228\pm0.010$  & $1710\pm100$   & $1.7E-15\pm5E-15$   & 13/8\\
                 & y=2.3          & --        &$0.062\pm0.005$  & $0.215\pm0.011$  & $1730\pm97$    & $1.4E-16\pm7E-17$   & 17/7\\
  \cline{2-8}
  Fig. 2         & y=0.1          & $\pi^-$   &$0.108\pm0.006$  & $0.437\pm0.008$  & $2000\pm104$   & $4.4E-6\pm6E-7$     & 1/-\\
  p-p            & y=0.3          & --        &$0.106\pm0.005$  & $0.457\pm0.009$  & $1800\pm100$   & $4.5E-7\pm5E-8$     & 1/2\\
  80 GeV         & y=0.5          &--         &$0.102\pm0.006$  & $0.349\pm0.012$  & $1780\pm72$    & $9E-8\pm5E-9$       & 1/2\\
                 & y=0.7          &--         &$0.097\pm0.006$  & $0.368\pm0.012$  & $1800\pm110$   & $8.5E-9\pm7E-10$    & 1/3\\
                 & y=0.9          & --        &$0.093\pm0.006$  & $0.367\pm0.010$  & $1820\pm90$    & $7.3E-10\pm4E-11$   & 4/4\\
                 & y=1.1          & --        &$0.088\pm0.004$  & $0.389\pm0.011$  & $1820\pm108$   & $6.3E-11\pm6E-12$   & 4/4\\
                 & y=1.3          & --        &$0.084\pm0.004$  & $0.235\pm0.008$  & $1760\pm100$   & $2.2E-11\pm6E-12$   & 17/6\\
                 & y=1.5          & --        &$0.080\pm0.005$  & $0.249\pm0.009$  & $1680\pm100$   & $2E-12\pm5E-13$     & 16/4\\
                 & y=1.7          & --        &$0.076\pm0.006$  & $0.260\pm0.010$  & $1735\pm90$    & $1.4E-13\pm4E-14$   & 16/3\\
                 & y=1.9          & --        &$0.072\pm0.006$  & $0.250\pm0.008$  & $1750\pm86$    & $1.1E-14\pm6E-15$   & 9/2\\
                 & y=2.1          & --        &$0.069\pm0.005$  & $0.220\pm0.009$  & $1680\pm80$    & $1E-15\pm5E-15$     & 14/2\\
                 & y=2.3          & --        &$0.064\pm0.005$  & $0.211\pm0.012$  & $1700\pm70$    & $7E-17\pm5E-18$     & 15/-\\
  \cline{2-8}
  Fig. 1         & y=0.1          & $\pi^+$   &$0.115\pm0.004$  & $0.425\pm0.008$  & $2250\pm100$   & $4.2E-6\pm5E-7$     & 0.3/1\\
  p-p            & y=0.3          & --        &$0.111\pm0.004$  & $0.420\pm0.009$  & $2150\pm70$    & $4.7E-7\pm6E-8$     & 0.4/2\\
  158 GeV        & y=0.5          &--         &$0.108\pm0.004$  & $0.430\pm0.012$  & $2160\pm96$    & $4.4E-8\pm5E-9$     & 1/3\\
                 & y=0.7          &--         &$0.104\pm0.005$  & $0.330\pm0.011$  & $2160\pm60$    & $9.2E-9\pm4E-10$    & 0.5/4\\
                 & y=0.9          & --        &$0.100\pm0.006$  & $0.300\pm0.012$  & $2200\pm110$   & $1E-9\pm6E-10$      & 4/4\\
                 & y=1.1          & --        &$0.095\pm0.005$  & $0.348\pm0.010$  & $2000\pm106$   & $9.5E-11\pm7E-12$   & 5/5\\
                 & y=1.3          & --        &$0.091\pm0.004$  & $0.360\pm0.010$  & $2000\pm122$   & $9E-12\pm6E-13$     & 16/5\\
                 & y=1.5          & --        &$0.087\pm0.005$  & $0.370\pm0.011$  & $1900\pm120$   & $9E-13\pm7E-14$     & 13/5\\
                 & y=1.7          & --        &$0.082\pm0.006$  & $0.370\pm0.008$  & $1900\pm108$   & $8.6E-14\pm6E-15$   & 21/5\\
                 & y=1.9          & --        &$0.078\pm0.004$  & $0.280\pm0.012$  & $1900\pm107$   & $7.3E-15\pm5E-16$   & 19/5\\
                 & y=2.1          & --        &$0.075\pm0.006$  & $0.423\pm0.007$  & $1990\pm120$   & $5.5E-16\pm5E-17$   & 26/5\\
                 & y=2.3          & --        &$0.070\pm0.005$  & $0.393\pm0.008$  & $2019\pm110$   & $5E-17\pm7E-18$     & 42/4\\
\hline
Fig. 2          & y=0.1           & $\pi^-$   &$0.114\pm0.004$  & $0.440\pm0.010$   & $2235\pm110$   & $3.9E-6\pm7E-7$     & 0.4/1\\
p-p             & y=0.3           & --        &$0.112\pm0.006$  & $0.435\pm0.012$  & $2100\pm80$    & $4.4E-7\pm7E-8$     & 0.1/2\\
  158 GeV        & y=0.5          &--         &$0.108\pm0.004$  & $0.430\pm0.012$  & $2160\pm96$    & $4E-8\pm5E-9$       & 1.5/3\\
                 & y=0.7          &--         &$0.105\pm0.006$  & $0.330\pm0.010$  & $2160\pm90$    & $8.4E-9\pm4E-10$    & 1/4\\
                 & y=0.9          & --        &$0.100\pm0.006$  & $0.330\pm0.012$  & $2200\pm100$   & $8E-10\pm5E-11$     & 1/7\\
                 & y=1.1          & --        &$0.096\pm0.005$  & $0.352\pm0.011$  & $2010\pm91$    & $7E-11\pm4E-12$     & 2/7\\
                 & y=1.3          & --        &$0.092\pm0.005$  & $0.366\pm0.010$  & $2120\pm102$   & $6.4E-12\pm4E-13$   & 3/6\\
                 & y=1.5          & --        &$0.087\pm0.006$  & $0.370\pm0.010$  & $1900\pm100$   & $6E-13\pm5E-15$     & 3.5/7\\
                 & y=1.7          & --        &$0.083\pm0.005$  & $0.370\pm0.012$  & $1909\pm98$    & $5.6E-14\pm5E-15$   & 5/6\\
                 & y=1.9          & --        &$0.080\pm0.004$  & $0.400\pm0.010$  & $1900\pm100$   & $4E-15\pm7E-16$     & 2/5\\
                 & y=2.1          & --        &$0.076\pm0.005$  & $0.423\pm0.013$  & $1988\pm110$   & $3.2E-16\pm5E-17$   & 1/3\\
                 & y=2.3          & --        &$0.072\pm0.005$  & $0.398\pm0.013$  & $2020\pm110$   & $2E-17\pm5E-18$   & 1/3\\
 \hline
Fig. 3          & y=0.1         & $K^+$       &$0.091\pm0.005$   & $0.250\pm0.010$  & $700\pm85$    & $1.5E-6\pm5E-7$     & 6/3\\
 p-p            & y=0.3          & --         &$0.087\pm0.005$  & $0.200\pm0.010$   & $700\pm62$    & $1.7E-7\pm4E-8$     & 5/5\\
 20 GeV         & y=0.5          &--          &$0.082\pm0.005$  & $0.280\pm0.008$   & $720\pm66$    & $1.2E-8\pm6E-9$     & 7/2\\
                & y=0.7          & --         &$0.079\pm0.006$  & $0.210\pm0.012$  & $665\pm88$     & $1.45E-9\pm5E-10$   & 27/6\\
                & y=0.9          & --         &$0.076\pm0.005$  & $0.250\pm0.009$  & $665\pm70$     & $1E-10\pm6E-11$     & 5/8\\
                & y=1.1          & --         &$0.074\pm0.006$  & $0.248\pm0.009$  & $600\pm60$     & $6E-12\pm6E-13$     & 15/7\\
                & y=1.3          & --         &$0.069\pm0.004$  & $0.245\pm0.009$  & $600\pm71$     & $5E-13\pm7E-14$     & 8/4\\
                & y=1.5          & --         &$0.066\pm0.006$  & $0.219\pm0.011$  & $635\pm80$     & $3.9E-14\pm7E-15$   & 12/4\\
\hline
\end{tabular}%
\end{center}}
\end{table*}

\begin{table*}
{\scriptsize Table 1. Continue.
\vspace{-.50cm}
\begin{center}
\begin{tabular}{cccccccccc}\\ \hline\hline
Collisions       & Rapidity       & Particle  & $T_0$ (GeV)    & $\beta_T$ (c)    & $V (fm^3)$     & $N_0$      & $\chi^2$/ dof \\ \hline
Fig. 4          & y=0.1           & $K^-$     &$0.091\pm0.006$  & $0.260\pm0.010$  & $700\pm92$    & $6.1E-7\pm4E-8$     & 10/6\\
 p-p            & y=0.3          & --         &$0.087\pm0.005$  & $0.200\pm0.010$  & $700\pm62$    & $5.4E-8\pm5E-9$     & 6/5\\
 20 GeV         & y=0.5          &--          &$0.082\pm0.005$  & $0.280\pm0.008$  & $722\pm53$    & $4.8E-9\pm5E-10$    & 3/2\\
                & y=0.7          & --         &$0.079\pm0.005$  & $0.200\pm0.011$  & $680\pm70$    & $3.3E-10\pm7E-11$   & 8/2\\
                & y=0.9          & --         &$0.076\pm0.004$  & $0.250\pm0.007$  & $665\pm55$    & $2.2E-11\pm5E-2$    & 5/4\\
                & y=1.1          & --         &$0.072\pm0.005$  & $0.250\pm0.008$  & $605\pm50$    & $1.3E-12\pm5E-13$   & 8/2\\
                & y=1.3          & --         &$0.068\pm0.005$  & $0.245\pm0.008$  & $605\pm65$    & $8.5E-14\pm5E-15$   & 2/-\\
                & y=1.5          & --         &$0.064\pm0.005$  & $0.225\pm0.009$  & $630\pm70$    & $4.6E-15\pm4E-16$   & 2/-\\
  \cline{2-8}
 Fig. 3         & y=0.1         & $K^+$       &$0.101\pm0.006$  & $0.310\pm0.010$  & $1004\pm120$  & $1.6E-6\pm6E-7$    & 1.5/5\\
   p-p          & y=0.3          & --         &$0.097\pm0.006$  & $0.298\pm0.013$ & $1000\pm95$    & $1.6E-7\pm4E-8$     & 1/7\\
  31 GeV        & y=0.5          &--          &$0.094\pm0.006$  & $0.340\pm0.010$ & $900\pm70$     & $1.6E-8\pm5E-9$     & 0.2/4\\
                & y=0.7          & --         &$0.091\pm0.004$  & $0.330\pm0.010$ & $900\pm90$     & $1.4E-9\pm6E-10$    & 4.5/7\\
                & y=0.9          &--          &$0.086\pm0.005$  & $0.323\pm0.012$ & $830\pm110$    & $1.3E-10\pm5E-11$   & 5/7\\
                & y=1.1          & --         &$0.082\pm0.004$  & $0.310\pm0.008$ & $830\pm70$     & $3.2E-11\pm4E-12$   & 15/7\\
                & y=1.3          &--          &$0.079\pm0.005$  & $0.320\pm0.012$ & $770\pm85$     & $1.2E-12\pm5E-13$   & 12/7\\
                & y=1.5          & --         &$0.075\pm0.006$  & $0.330\pm0.009$ & $780\pm76$     & $9.6E-14\pm5E-15$   & 9/7\\
                & y=1.7          &--          &$0.072\pm0.005$  & $0.270\pm0.011$ & $720\pm50$     & $8E-15\pm7E-16$     & 9/6\\
                & y=1.9          & --         &$0.067\pm0.005$  & $0.270\pm0.012$ & $750\pm78$     & $5E-16\pm5E-17$     & 4.5/5\\
                & y=2.1          &--          &$0.068\pm0.006$  & $0.230\pm0.010$ & $700\pm80$     & $4E-17\pm4E-18$     & 33/4\\
                & y=2.3          & --         &$0.063\pm0.004$  & $0.200\pm0.008$ & $700\pm75$     & $3.6E-18\pm5E-19$   & 1/1\\
\cline{2-8}
  Fig. 4        & y=0.1          & $K^-$      &$0.100\pm0.004$  & $0.310\pm0.009$ & $1000\pm100$   & $6E-7\pm4E-8$       & 1.5/3\\
   p-p          & y=0.3          & --         &$0.096\pm0.006$  & $0.299\pm0.013$ & $1000\pm95$    & $5.5E-8\pm4E-9$     & 5/4\\
  31 GeV        & y=0.5          &--          &$0.093\pm0.005$  & $0.340\pm0.011$ & $900\pm70$     & $5E-9\pm5E-10$      & 0.4/2\\
                & y=0.7          & --         &$0.090\pm0.006$  & $0.260\pm0.013$ & $900\pm100$    & $4.4E-10\pm4E-11$   & 2/4\\
                & y=0.9          &--          &$0.085\pm0.006$  & $0.320\pm0.010$ & $830\pm110$    & $4E-11\pm5E-12$     & 4.5/4\\
                & y=1.1          & --        &$0.082\pm0.006$  & $0.310\pm0.012$ & $830\pm90$     & $2.9E-12\pm7E-13$   & 6/3\\
                & y=1.3          &--         &$0.078\pm0.005$  & $0.318\pm0.010$ & $780\pm80$     & $1.8E-13\pm5E-14$   & 1/3\\
                & y=1.5          & --        &$0.074\pm0.005$  & $0.330\pm0.013$ & $780\pm80$     & $1.2E-14\pm5E-15$   & 0.4/1\\
                & y=1.7          &--         &$0.070\pm0.004$  & $0.210\pm0.012$ & $750\pm70$     & $5E-16\pm5E-17$     & 5/-\\
                & y=1.9          & --         &$0.067\pm0.006$  & $0.270\pm0.012$ & $750\pm80$     & $3.1E-17\pm6E-18$   & 0.2/-\\
\cline{2-8}
   Fig. 3       & y=0.1        & $K^+$        &$0.114\pm0.005$  & $0.233\pm0.010$ & $1345\pm100$  & $1.25E-6\pm5E-7$     & 6/7\\
   p-p          & y=0.3          & --         &$0.111\pm0.006$  & $0.240\pm0.010$ & $1300\pm80$    & $1.3E-7\pm6E-8$     & 1.5/6\\
   40 GeV       & y=0.5          &--          &$0.108\pm0.006$  & $0.270\pm0.012$ & $1200\pm100$   & $1.1E-8\pm4E-9$     & 9/7\\
                & y=0.7          & --         &$0.103\pm0.006$  & $0.290\pm0.010$ & $1188\pm100$   & $1E-9\pm7E-10$      & 3/7\\
                & y=0.9          &--         &$0.099\pm0.004$  & $0.320\pm0.012$ & $1260\pm110$   & $9.4E-11\pm4E-12$   & 3/7\\
                & y=1.1          & --        &$0.096\pm0.005$  & $0.350\pm0.010$ & $1230\pm100$   & $8.5E-12\pm7E-13$   & 4/7\\
                & y=1.3          &--         &$0.093\pm0.005$  & $0.330\pm0.009$ & $1200\pm100$   & $7.5E-13\pm6E-14$   & 5/7\\
                & y=1.5          & --       &$0.089\pm0.004$  & $0.320\pm0.010$ & $1200\pm80$    & $6E-14\pm5E-15$     & 6/7\\
                & y=1.7          &--         &$0.084\pm0.004$  & $0.208\pm0.010$ & $1110\pm104$   & $1.6E-14\pm6E-15$   & 11/7\\
                & y=1.9          & --        &$0.080\pm0.005$  & $0.290\pm0.012$ & $1200\pm108$   & $4.45E-16\pm4E-17$  & 19/6\\
\cline{2-8}
   Fig. 4       & y=0.1        & $K^-$       &$0.112\pm0.004$  & $0.240\pm0.012$  & $1350\pm127$  & $6E-7\pm4E-8$       & 2/5\\
   p-p          & y=0.3          & --        &$0.110\pm0.005$  & $0.240\pm0.010$ & $1300\pm87$    & $5.7E-8\pm5E-9$     & 1/4\\
   40 GeV       & y=0.5          &--         &$0.106\pm0.005$  & $0.260\pm0.008$ & $1200\pm109$   & $5E-9\pm7E-10$      & 11/5\\
                & y=0.7          & --        &$0.102\pm0.005$  & $0.285\pm0.009$ & $1200\pm110$   & $4E-10\pm6E-11$     & 3/5\\
                & y=0.9          &--         &$0.098\pm0.006$  & $0.310\pm0.011$ & $1260\pm120$   & $3.2E-11\pm6E-12$   & 1/4\\
                & y=1.1          & --        &$0.095\pm0.005$  & $0.350\pm0.009$ & $1235\pm100$   & $2.7E-12\pm4E-13$   & 2/4\\
                & y=1.3          &--         &$0.092\pm0.005$  & $0.280\pm0.010$ & $1205\pm87$    & $1.6E-13\pm5E-14$   & 2/3\\
                & y=1.5          & --        &$0.088\pm0.006$  & $0.250\pm0.009$ & $1200\pm94$    & $9.8E-15\pm6E-16$   & 2/2\\
                & y=1.7          &--         &$0.084\pm0.004$  & $0.208\pm0.012$ & $1100\pm150$   & $7E-16\pm5E-17$     & 5/2\\
                & y=1.9          & --        &$0.080\pm0.005$  & $0.290\pm0.012$ & $1200\pm108$   & $1.25E-17\pm5E-18$  & 2.5/-\\
                & y=2.1          & --        &$0.076\pm0.005$  & $0.200\pm0.010$ & $1200\pm100$   & $1.7E-18\pm4E-19$   & 0.1/-\\
\cline{2-8}
   Fig. 3         & y=0.1        & $K^+$     &$0.121\pm0.005$  & $0.240\pm0.010$ & $1708\pm93$    & $1.1E-6\pm6E-7$     & 1.5/5\\
   p-p          & y=0.3          & --       &$0.118\pm0.006$  & $0.240\pm0.007$ & $1600\pm68$    & $1.2E-7\pm6E-8$     & 2/7\\
   80 GeV       & y=0.5          &--        &$0.114\pm0.005$  & $0.280\pm0.011$ & $1660\pm120$   & $1.1E-8\pm4E-9$     & 1.5/7\\
                & y=0.7          & --       &$0.110\pm0.005$  & $0.320\pm0.010$ & $1630\pm120$   & $1E-9\pm6E-10$      & 1.5/6\\
                & y=0.9          &--        &$0.107\pm0.005$  & $0.290\pm0.012$ & $1638\pm142$   & $9E-11\pm4E-12$     & 2/5\\
                & y=1.1          & --        &$0.105\pm0.006$  & $0.300\pm0.009$ & $1601\pm60$    & $8E-12\pm6E-13$     & 2/7\\
                & y=1.3          &--         &$0.100\pm0.006$  & $0.272\pm0.008$ & $1601\pm90$    & $6E-14\pm7E-15$     & 8/7\\
                & y=1.5          & --       &$0.096\pm0.005$  & $0.305\pm0.010$  & $1578\pm90$    & $5E-15\pm5E-16$    & 2/7\\
                & y=1.7          &--         &$0.093\pm0.006$  & $0.300\pm0.010$ & $1500\pm100$   & $5.1E-15\pm5E-16$   & 4/6\\
                & y=1.9          & --        &$0.090\pm0.005$  & $0.250\pm0.011$ & $1520\pm108$   & $3.5E-16\pm6E-17$   & 4/6\\
\hline
\end{tabular}%
\end{center}}
\end{table*}

\begin{table*}
{\scriptsize Table 1. Continue.
\vspace{-.50cm}
\begin{center}
\begin{tabular}{cccccccccc}\\ \hline\hline
Collisions      & Rapidity       & Particle  & $T_0$ (GeV)    & $\beta_T$ (c)   & $V (fm^3)$     & $N_0$      & $\chi^2$/ dof \\ \hline
 Fig. 4         & y=0.1        & $K^-$       &$0.121\pm0.006$  & $0.240\pm0.010$ & $1700\pm115$   & $5.3E-7\pm4E-8$     & 1.5/5\\
   p-p          & y=0.3          & --        &$0.118\pm0.006$  & $0.240\pm0.007$ & $1600\pm79$    & $6E-8\pm4E-9$       & 1/4\\
   80 GeV       & y=0.5          &--         &$0.114\pm0.004$  & $0.280\pm0.011$ & $1660\pm123$   & $5.8E-9\pm6E-10$    & 2/4\\
                & y=0.7          & --        &$0.110\pm0.005$  & $0.314\pm0.010$ & $1630\pm120$   & $5E-10\pm5E-11$     & 0.4/4\\
                & y=0.9          &--         &$0.107\pm0.005$  & $0.270\pm0.010$ & $1638\pm142$   & $3.8E-11\pm6E-12$   & 2/4\\
                & y=1.1          & --        &$0.103\pm0.004$  & $0.290\pm0.008$ & $1621\pm120$   & $2.9E-12\pm5E-13$   & 16/5\\
                & y=1.3          &--         &$0.100\pm0.004$  & $0.262\pm0.012$ & $1607\pm95$    & $2.3E-13\pm5E-14$   & 5/5\\
                & y=1.5          & --        &$0.096\pm0.005$  & $0.300\pm0.010$ & $1578\pm98$    & $1.7E-14\pm5E-15$   & 0.7/2\\
                & y=1.7          &--         &$0.092\pm0.005$  & $0.270\pm0.010$ & $1535\pm110$   & $1.2E-15\pm7E-16$   & 6.5/\\
                & y=1.9          & --        &$0.089\pm0.004$  & $0.250\pm0.010$ & $1543\pm100$   & $5.6E-17\pm5E-18$   & 2/-\\
                & y=2.1          & --         &$0.085\pm0.006$  & $0.200\pm0.012$ & $1570\pm100$   & $6E-19\pm6E-20$     & 0.2/-\\
\cline{2-8}
Fig. 3          & y=0.1          & $K^+$      &$0.128\pm0.006$  & $0.230\pm0.008$ & $2130\pm120$   & $9E-7\pm7E-8$       & 2/6\\
   p-p          & y=0.3          & --         &$0.126\pm0.005$  & $0.250\pm0.011$ & $2117\pm110$   & $9.1E-8\pm7E-9$     & 1/6\\
   158 GeV      & y=0.5          &--          &$0.121\pm0.005$  & $0.275\pm0.012$ & $2110\pm110$   & $9.2E-9\pm5E-10$    & 1/6\\
                & y=0.7          & --         &$0.120\pm0.005$  & $0.295\pm0.011$ & $2025\pm120$   & $9E-10\pm6E-11$     & 0.5/6\\
                & y=0.9          &--          &$0.116\pm0.005$  & $0.280\pm0.012$ & $2000\pm115$   & $8.5E-11\pm5E-12$   & 1/6\\
                & y=1.1          & --         &$0.112\pm0.005$  & $0.240\pm0.010$ & $2070\pm130$   & $7E-12\pm6E-13$     & 2/6\\
                & y=1.3          &--          &$0.107\pm0.004$  & $0.278\pm0.011$ & $2050\pm112$   & $6E-13\pm6E-14$     & 1/5\\
                & y=1.5          & --         &$0.105\pm0.004$  & $0.248\pm0.011$ & $2020\pm103$   & $5E-14\pm7E-15$     & 2/5\\
                & y=1.7          &--          &$0.103\pm0.005$  & $0.268\pm0.010$ & $1980\pm120$   & $4.1E-15\pm7E-16$   & 2/5\\
                & y=1.9          & --         &$0.098\pm0.006$  & $0.217\pm0.010$ & $1966\pm106$   & $3E-16\pm4E-17$     & 5.5/4\\
                & y=2.1          & --         &$0.097\pm0.006$  & $0.210\pm0.011$ & $1993\pm100$   & $2E-17\pm5E-18$     & 11/5\\
                & y=2.3          & --         &$0.094\pm0.005$  & $0.300\pm0.009$ & $1960\pm100$   & $4E-18\pm5E-19$     & 13/3\\
                & y=2.5          & --         &$0.090\pm0.004$  & $0.230\pm0.010$ & $1976\pm90$    & $2E-19\pm4E-20$     & 1.5/-\\
\cline{2-8}
Fig. 4          & y=0.1          & $K^-$      &$0.128\pm0.005$  & $0.230\pm0.009$ & $2130\pm126$   & $6E-7\pm6E-8$       & 1/6\\
   p-p          & y=0.3          & --         &$0.125\pm0.004$  & $0.255\pm0.007$ & $2130\pm100$   & $5.8E-8\pm6E-9$     & 0.3/5\\
   158 GeV      & y=0.5          &--          &$0.122\pm0.006$  & $0.275\pm0.010$ & $2130\pm103$   & $6E-9\pm6E-9$       & 0.5/6\\
                & y=0.7          & --         &$0.118\pm0.004$  & $0.295\pm0.012$ & $2057\pm102$   & $5.4E-10\pm4E-11$   & 1/5\\
                & y=0.9          &--          &$0.114\pm0.006$  & $0.280\pm0.012$ & $2000\pm115$   & $4.5E-11\pm7E-12$   & 1/7\\
                & y=1.1          & --         &$0.110\pm0.006$  & $0.230\pm0.008$ & $2100\pm117$   & $3.7E-12\pm6E-13$   & 3/7\\
                & y=1.3          & --         &$0.106\pm0.006$  & $0.270\pm0.008$ & $2070\pm117$   & $3E-13\pm5E-14$     & 11/6\\
                & y=1.5          &--          &$0.104\pm0.005$  & $0.238\pm0.008$ & $2020\pm105$   & $2.3E-14\pm5E-15$   & 2/6\\
                & y=1.7          & --         &$0.101\pm0.004$  & $0.268\pm0.011$ & $1987\pm88$    & $1.7E-15\pm6E-16$   & 30/4\\
                & y=1.9          &--          &$0.097\pm0.006$  & $0.217\pm0.009$ & $1971\pm90$    & $1E-16\pm5E-17$     & 1/3\\
                & y=2.1          & --         &$0.094\pm0.005$  & $0.210\pm0.009$ & $1993\pm110$   & $7E-18\pm5E-19$     & 37/2\\
                & y=2.3          & --         &$0.092\pm0.005$  & $0.290\pm0.010$ & $1570\pm100$   & $4.1E-19\pm4E-20$   & 2/1\\
                & y=2.5          & --         &$0.089\pm0.004$  & $0.230\pm0.008$ & $1980\pm120$   & $6.8E-20\pm6E-21$   & 4/-\\
\cline{2-8}
Fig. 5          & y=0.1          & $\bar p$   &$0.120\pm0.006$  & $0.089\pm0.009$ & $800\pm56$     & $1.01E-7\pm6E-8$    & 8/5\\
   p-p          & y=0.3          & --         &$0.117\pm0.005$  & $0.100\pm0.008$ & $800\pm70$     & $1.25E-8\pm4E-9$    & 20/5\\
   31 GeV       & y=0.5          &--          &$0.113\pm0.004$  & $0.060\pm0.008$ & $800\pm93$     & $6.3E-10\pm5E-11$   & 10/4\\
                & y=0.7          & --         &$0.109\pm0.005$  & $0.120\pm0.010$ & $700\pm68$     & $5.6E-11\pm4E-12$   & 12/3\\
                & y=0.9          &--          &$0.060\pm0.007$  & $0.120\pm0.007$ & $755\pm110$    & $2.6E-12\pm4E-13$   & 8/3\\
\cline{2-8}
Fig. 5          & y=0.1          & $\bar p$  &$0.129\pm0.004$  & $0.055\pm0.008$ & $1100\pm120$   & $8.4E-8\pm4E-9$     & 3/1\\
   p-p          & y=0.3          & --        &$0.125\pm0.006$  & $0.150\pm0.009$ & $1000\pm110$   & $9.8E-9\pm6E-10$    & 22/5\\
   40 GeV       & y=0.5          &--         &$0.121\pm0.005$  & $0.120\pm0.009$ & $1060\pm88$    & $7.8E-10\pm5E-11$   & 11/3\\
                & y=0.7          & --        &$0.116\pm0.004$  & $0.140\pm0.009$ & $1040\pm80$    & $5E-11\pm5E-12$     & 19/4\\
                & y=0.9          &--         &$0.090\pm0.005$  & $0.100\pm0.008$ & $1000\pm80$    & $2.5E-12\pm5E-13$   & 7/4\\
\cline{2-8}
Fig. 5          & y=0.1          & $\bar p$  &$0.137\pm0.006$  & $0.120\pm0.007$ & $1500\pm160$   & $1.5E-7\pm6E-8$     & 36/7\\
   p-p          & y=0.3          & --        &$0.133\pm0.005$  & $0.120\pm0.007$ & $1400\pm100$   & $1.5E-8\pm6E-9$     & 10/7\\
   80 GeV       & y=0.5          &--         &$0.124\pm0.006$  & $0.060\pm0.007$ & $1426\pm93$    & $1.25E-9\pm5E-10$   & 5/7\\
                & y=0.7          & --        &$0.122\pm0.005$  & $0.100\pm0.007$ & $1450\pm120$   & $1E-10\pm6E-11$     & 4/7\\
                & y=0.9          &--         &$0.110\pm0.005$  & $0.060\pm0.008$  & $1410\pm100$   & $8E-12\pm5E-13$     & 12/6\\
                & y=1.1          &--         &$0.107\pm0.005$  & $0.120\pm0.009$ & $1370\pm100$   & $4.8E-13\pm6E-14$   & 10/6\\
                & y=1.3          & --        &$0.102\pm0.006$  & $0.120\pm0.010$ & $1300\pm80$    & $3.3E-14\pm4E-15$   & 14/5\\
                & y=1.5          &--         &$0.086\pm0.006$  & $0.290\pm0.011$ & $1330\pm110$   & $6E-16\pm6E-17$     & 277/4\\
  \cline{2-8}
Fig. 5          & y=0.1          & $\bar p$  &$0.144\pm0.005$  & $0.060\pm0.006$ & $1850\pm110$   & $2.3E-7\pm4E-8$    & 4.5/8\\
   p-p          & y=0.3          & --         &$0.139\pm0.006$  & $0.100\pm0.007$ & $1730\pm120$   & $2.3E-8\pm5E-9$     & 107/8\\
158 GeV         & y=0.5          &--          &$0.134\pm0.005$  & $0.100\pm0.008$ & $1700\pm85$    & $2.2E-9\pm4E-10$    & 7/8\\
                & y=0.7          & --       &$0.130\pm0.006$  & $0.160\pm0.011$ & $1647\pm110$   & $2.2E-10\pm6E-11$   & 3/7\\
                & y=0.9          &--         &$0.121\pm0.006$  & $0.120\pm0.009$ & $1670\pm120$   & $1.65E-11\pm6E-12$  & 3/6\\
                & y=1.1          &--          &$0.117\pm0.004$  & $0.120\pm0.007$ & $1600\pm120$   & $1.3E-12\pm5E-13$   & 2/5\\
                & y=1.3          & --       &$0.113\pm0.005$  & $0.150\pm0.011$ & $1650\pm130$   & $1E-13\pm6E-14$     & 13/6\\
                & y=1.5          &--          &$0.110\pm0.005$  & $0.150\pm0.012$ & $1610\pm100$   & $5.3E-15\pm5E-16$   & 2/3\\
                & y=1.7          & --         &$0.106\pm0.004$  & $0.150\pm0.010$ & $1635\pm110$   & $3.3E-16\pm7E-17$   & 3/2\\
                & y=1.9          &--          &$0.088\pm0.005$  & $0.085\pm0.006$ & $1630\pm100$   & $1.1E-17\pm6E-18$   & 4/1\\
\hline
\end{tabular}%
\end{center}}
\end{table*}
legends in each panel. The spectra of $K^-$ at 0.5 and 0.7 rapidity slices are scaled with the factor 1/8
to avoid overlapping of the experimental data and solid curve with the others. The solid curve over the experimental
data shows our fitting result by the blast wave model with Boltzmann Gibbs statistics. It can be seen that the
Blast wave model provides an approximately well fitting of the data at all rapidity slices.

\begin{figure*}[htbp]
\begin{center}
\includegraphics[width=16.cm]{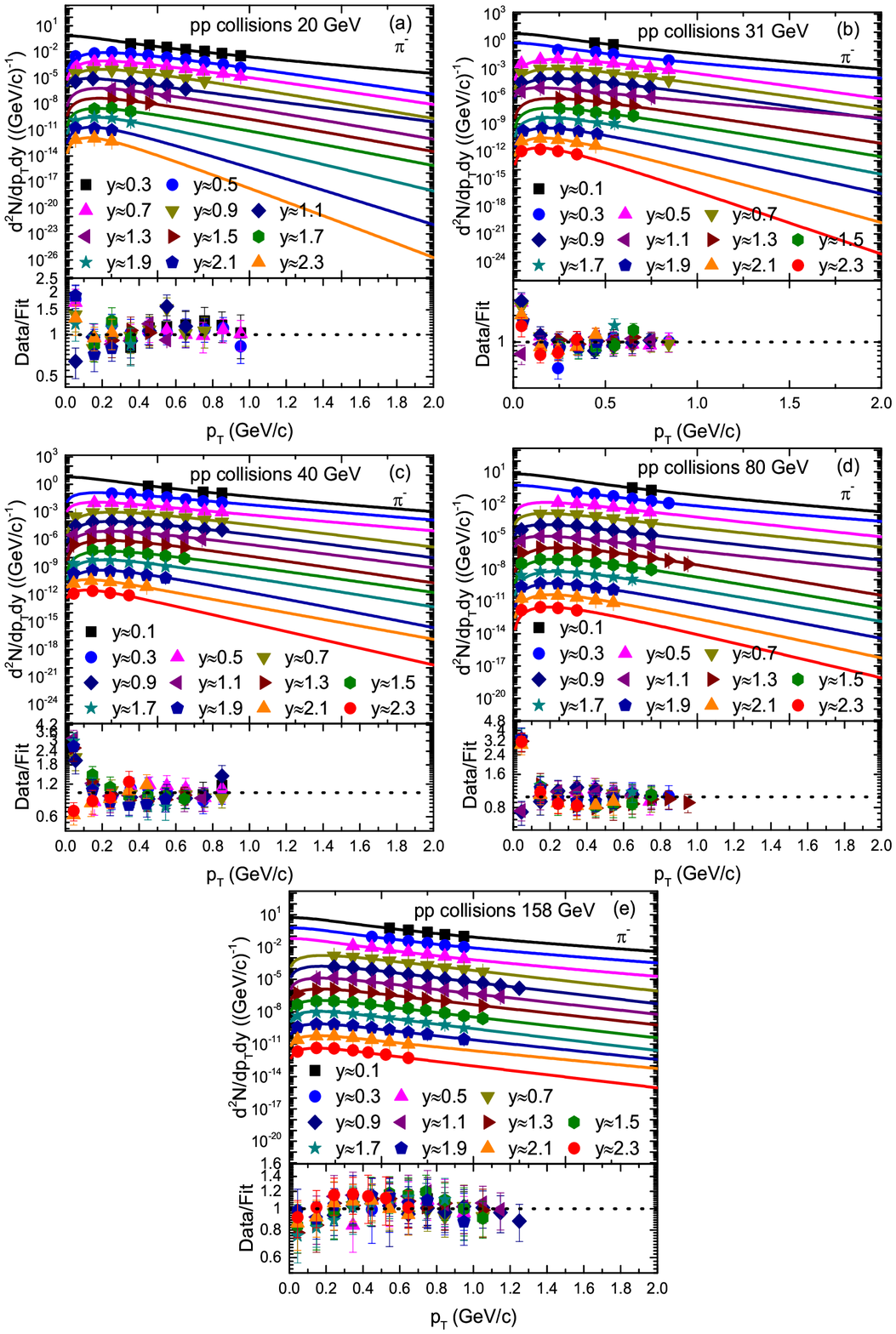}
\end{center}
Fig. 2. Transverse momentum spectra of $\pi^-$ produced in different rapidity
slices in pp collisions. Panel (a)-(e) corresponds to 20, 31, 40, 80 and 158 GeV energy respectively.
The symbols represent the experimental data of NA61/SHINE Collaboration measured at CERN [33]. The
curves are the results of out fits by the Blast Wave model with Boltzmann Gibs
statistics. The corresponding data/fit ratios are are followed in each panel.
\end{figure*}

\begin{figure*}[htbp]
\begin{center}
\includegraphics[width=16.cm]{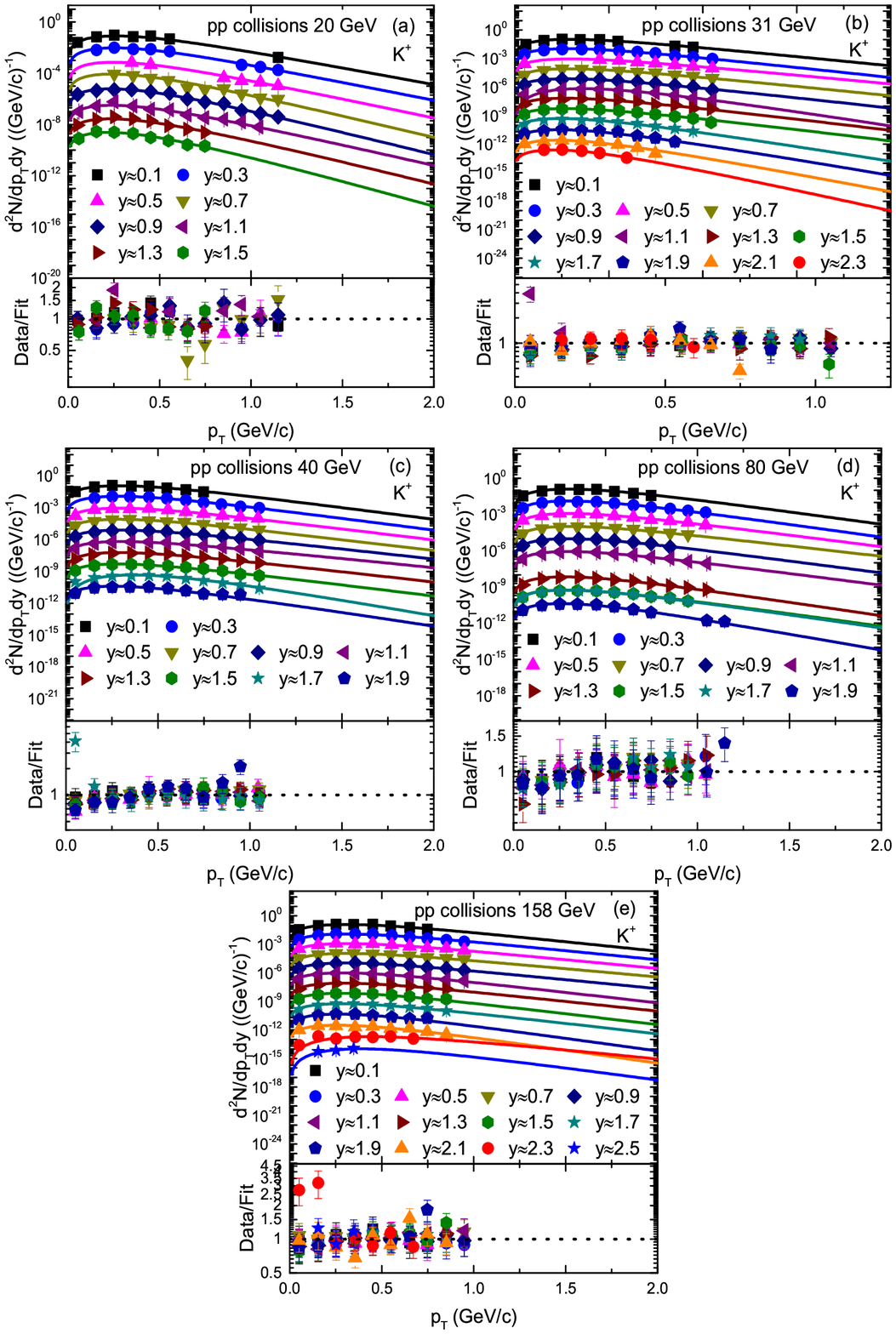}
\end{center}
Fig. 3. Transverse momentum spectra of $K^+$ produced in different rapidity
slices in pp collisions. Panel (a)-(e) corresponds to 20, 31, 40, 80 and 158 GeV energy respectively.
The symbols represent the experimental data of NA61/SHINE Collaboration measured at CERN [33]. The
curves are the results of out fits by the Blast Wave model with Boltzmann Gibs
statistics. The corresponding data/fit ratios are are followed in each panel.
\end{figure*}

Figure 5 is similar to fig. 4, but it shows the $p_T$ spectra of $\bar p$ in pp collisions. The symbols
denote the experimental data of NA61/SHINE Collaboration measured at CERN [33].
Different panels in fig. 5 show different collision energies, and different symbols represent
different rapidity slices. The solid curves are the results of our fit by Eq. (1) with fluctuations.

\begin{figure*}[htbp]
\begin{center}
\includegraphics[width=16.cm]{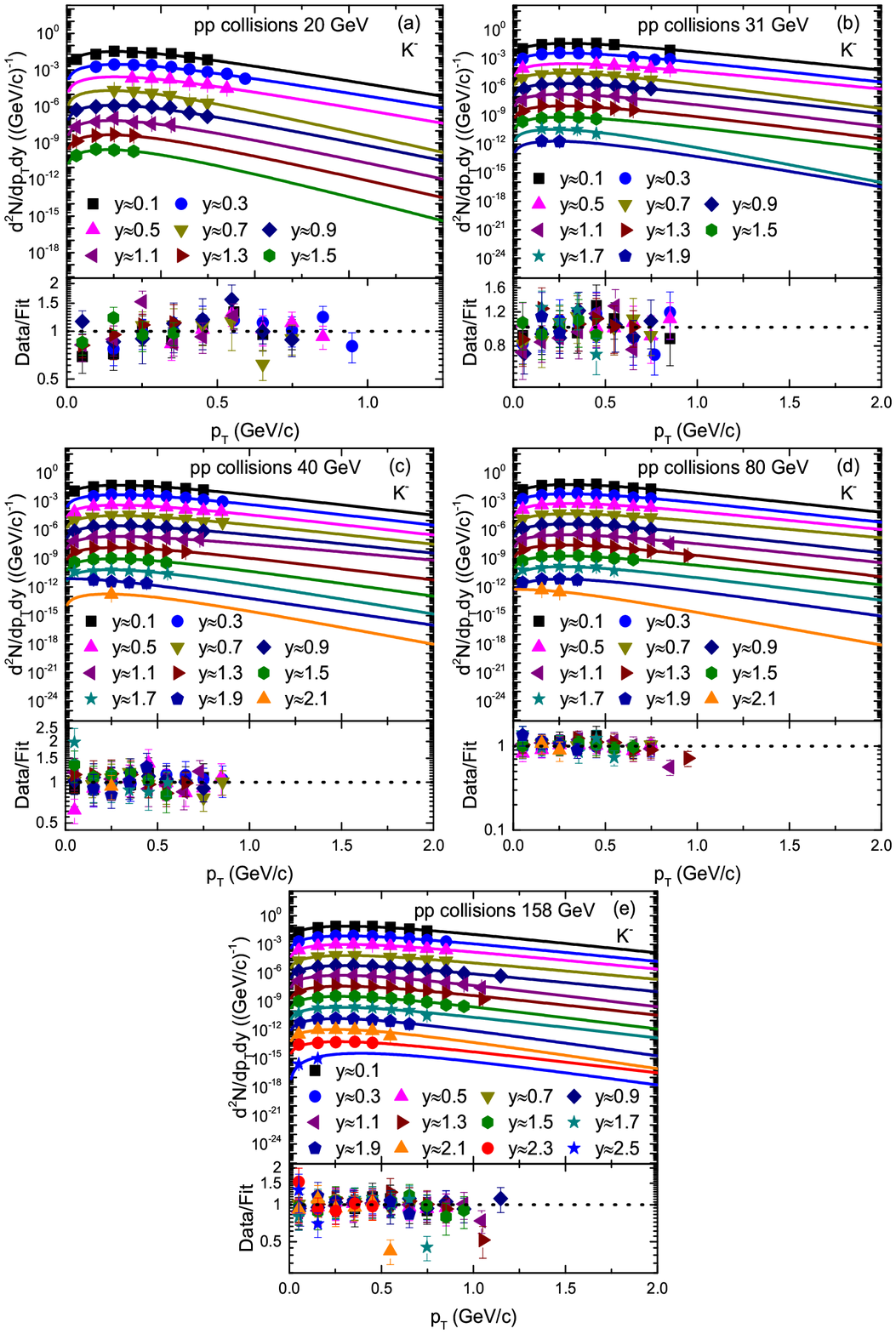}
\end{center}
Fig. 4. Transverse momentum spectra of $K^-$ produced in different rapidity
slices in pp collisions. Panel (a)-(e) corresponds to 20, 31, 40, 80 and 158 GeV energy respectively.
The symbols represent the experimental data of NA61/SHINE Collaboration measured at CERN [33]. The
curves are the results of out fits by the Blast Wave model with Boltzmann Gibs
statistics. The corresponding data/fit ratios are are followed in each panel.
\end{figure*}

We used the least square method in the fit process to get the minimum $\chi^2$. From fig.1-fig.5, the $\chi^2$
is large in some cases which shows that the dispersion between the curve and data is large, however the fitting is
approximately acceptable, but in most cases the model results describe the experimental data well in the $p_T$
spectra of the particles produced in different rapidity slices. Each panel in each figure is followed by the
corresponding result of its data/fit in order to show the dispersion of the curve from the data. In fact, the data/fit
ratio is large in some cases due to the large dispersion between the curve and data.

 It should be noted that the data used in this work is from a fixed target experiment, where energies are in lab frame.
Therefore we need to convert it to the center of mass energies.  The corresponding 20, 31, 40, 80 and 158 GeV/c energies in
the lab frame are equal to 6.3, 7.7, 8.8, 12.3 and 17.3 GeV respectively in the center of mass frame. In addition, The rapidity
of the particle was measured in the cms system as , $y$ = a tanh$\beta_L$,
where $\beta_L$ represents the longitudinal component of the velocity and is given by $\beta_L$= $p_L$/E  with c = 1,
whereas E and $p_L$ are the energy and longitudinal momentum in the cms frame.

\begin{figure*}[htbp]
\begin{center}
\includegraphics[width=16.cm]{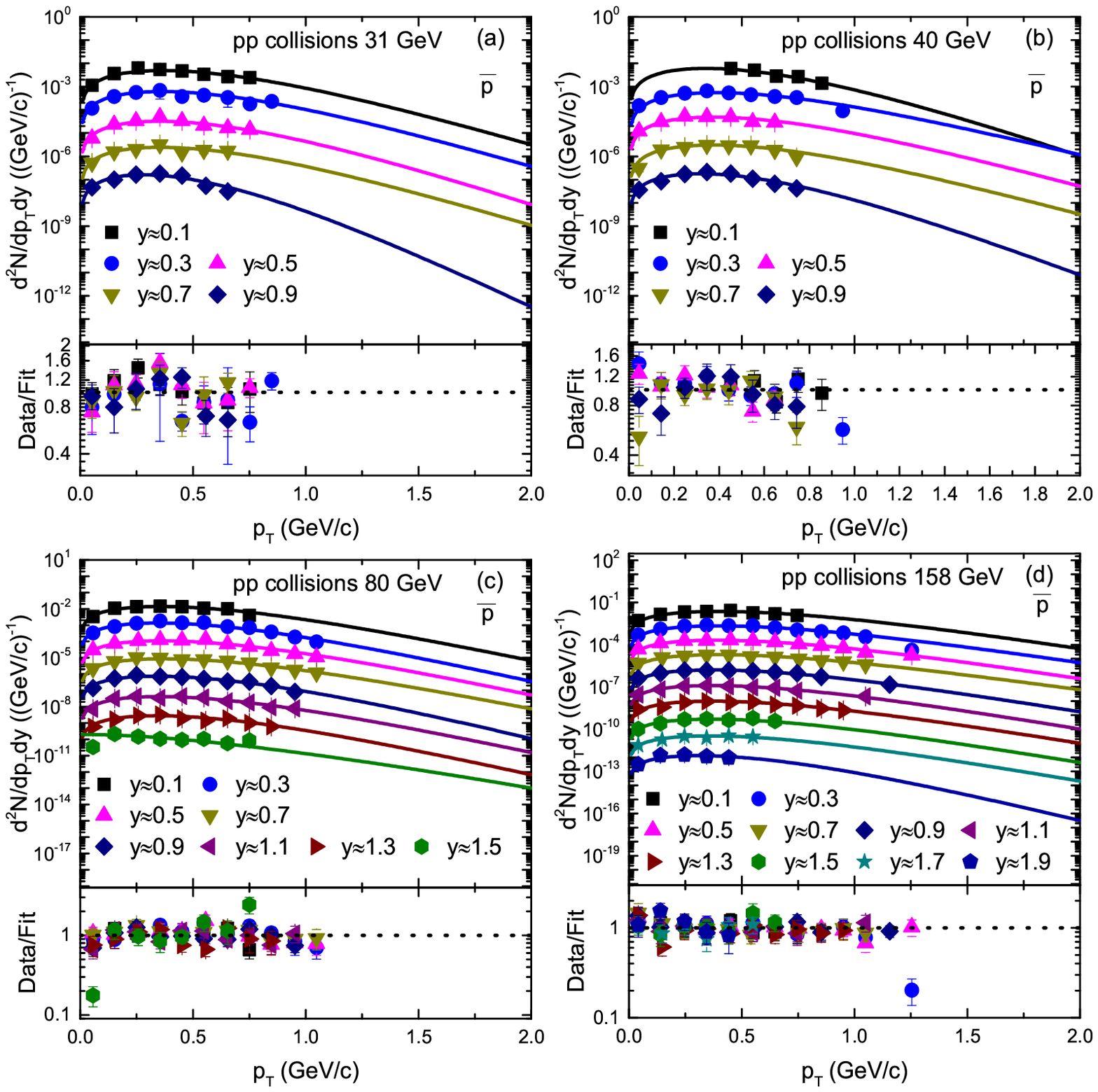}
\end{center}
Fig. 5. Transverse momentum spectra of $\bar p$ produced in different rapidity
slices in pp collisions. Panel (a)-(d) corresponds to 20, 31, 40, 80 and 158 GeV energy respectively.
The symbols represent the experimental data of NA61/SHINE Collaboration measured at CERN [33]. The
curves are the results of out fits by the Blast Wave model with Boltzmann Gibs
statistics. The corresponding data/fit ratios are are followed in each panel.
\end{figure*}

To study the change in trend of parameters with rapidity and collision energy, Figure 6 shows
the dependencies of kinetic freeze-out temperature ($T_0$) on rapidity and energy for the production of different
particles in pp collisions. Panel (a), (b) and (c) corresponds to pion, kaon and anti-proton respectively.
The closed and open symbols in panel (a)-(c) represent the positively and negatively charged particles respectively. The
trend of symbols from left to right shows the dependence of kinetic freeze-out temperature ($T_0$) on rapidity.
While the dependence of $T_0$ on energy is shown by the symbols from up to downward. One can see that $T_0$
increases with the increases of collision energy. The reason behind this is that when the energy
increases, the collision becomes more violent and transfers more energy which results in higher excitation of
the system and naturally the system with high degree of excitation has high $T_0$. On the other hand, the kinetic
freeze-out temperature ($T_0$) decreases with the increase of rapidity from mid-rapidity region to forward
rapidity, because when rapidity increases, the energy transfer in the system decreases due to large penetration
between participant nucleons. In the meantime, due to less produced particles taking part in the scattering
process, these degree of multiple scattering also decreases, the $p_T$ decreases due to both the factors. In addition,
it is observed that $T_0$ increases for heavier particles which shows the mass differential kinetic freeze-out scenario.

\begin{figure*}[htbp]
\begin{center}
\includegraphics[width=16.cm]{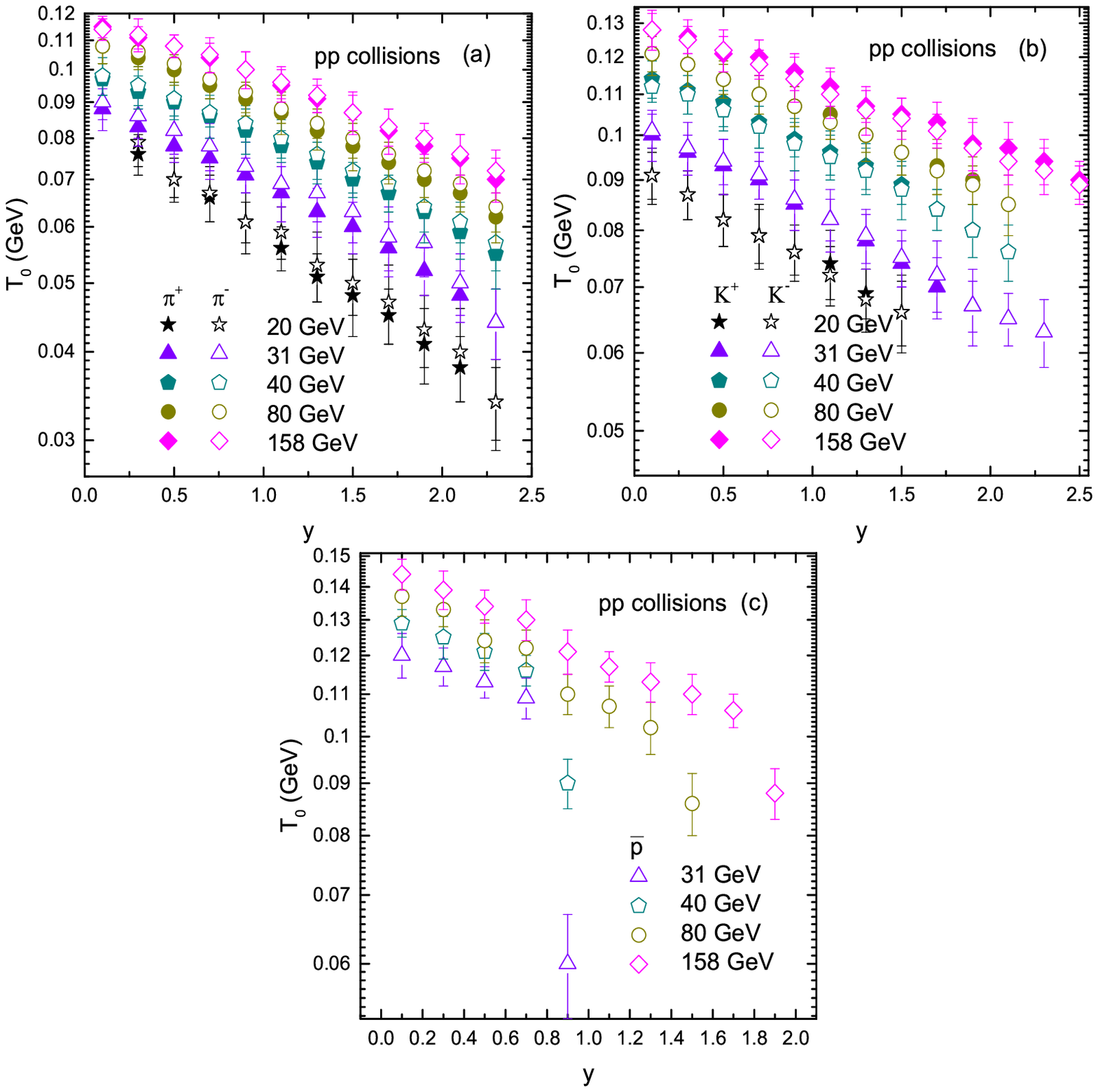}
\end{center}
Fig. 6. Dependence of $T_0$ on rapidity and collision energy.
\end{figure*}

Figure 7 is similar to fig.6, but it shows the dependence of $\beta_T$ on rapidity and energy. At present, we did not
observe any dependence of $\beta_T$ on rapidity and energy. Although there is energy dependence of $\beta_T$ in literature [10, 17],
but we can study it in more detail by analyzing more data for different particles in different collisions with different models.
Furthermore, $\beta_T$ is mass dependent and it is larger for lighter particles.

 We would like to point out that QGP like properties are reported at LHC energies in pp collisions, where the values of $T_0$ and
$\beta_T$ are reported to be $163\pm10$ MeV and  $0.49\pm0.02$ respectively [34]. It is also not possible to observe any such effect at the
energies under consideration but the relevant parameter can still be checked for low energies to understand the nature of the collisions
in the final state in comparison to high energy pp collisions. The values observed in our case are small compared to the one listed
above that shows that no such effect is observed in low energy pp collisions.

\begin{figure*}[htbp]
\begin{center}
\includegraphics[width=16.cm]{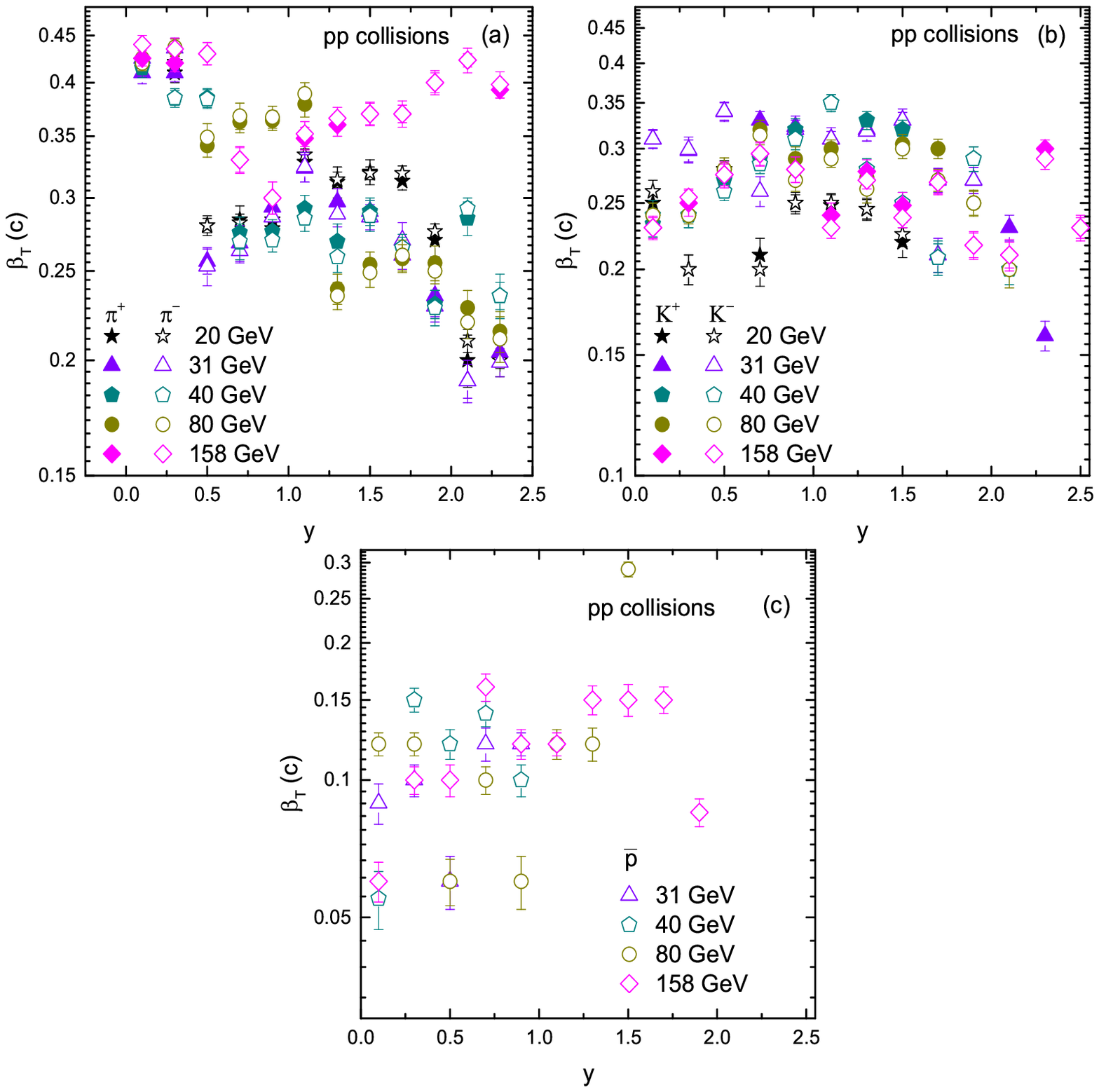}
\end{center}
Fig. 7. Dependence of $\beta_T$ on rapidity and collision energy.
\end{figure*}

Both $T_0$ and $\beta_T$ show mass dependency, which reflects the formation time dependence. Hydrodynamically, heavier particles coming out of
the system earlier in time with smaller radial flow velocities. This shows that as the mass increases, the formation time as well as
$\beta_T$ decreases whereas $T_0$ increases. Indeed, there are various hydrodynamical simulations which observe one set of parameters
for all the particles (common freeze-out temperature as well as transverse flow velocity), but their explanation is different. Besides, we can
get different results from different models, even from the same model we can get different results if the method is different and the limitations
and conditions are different. The selection of $T_0$ and $\beta_T$ is very technical and complex, such as in some cases, it depends on the range
of $p_T$ spectra, and it may also depend on the value of $n_0$. If the value of $n_0$ is 1 or 2, it does not have an effect on the results of
parameters but if it is taken as free parameter, then the free parameters (especially $\beta_T$) fluctuates obviously.

\begin{figure*}[htbp]
\begin{center}
\includegraphics[width=16.cm]{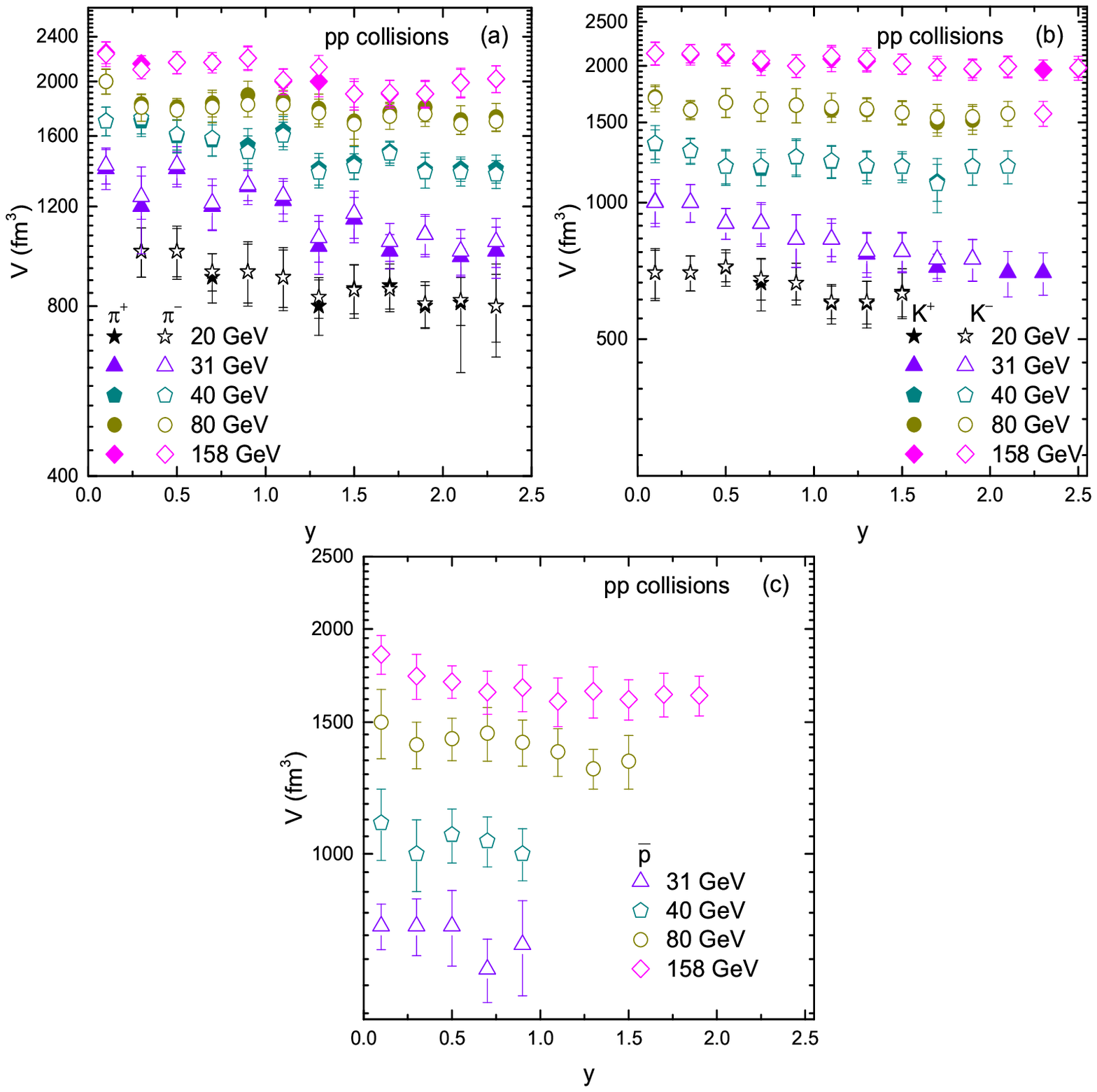}
\end{center}
Fig. 8. Dependence of $V$ on rapidity and collision energy.
\end{figure*}

Figure 8 is similar to fig. 6 and 7, but it shows the dependence of $V$ on rapidity and energy. One can see that the kinetic freeze-out volume
increases with energy. The reason behind this is that larger initial bulk system exists at high energy. The evolution time becomes longer
at higher energies, and it corresponds to larger partonic system and naturally the kinetic freeze-out volume becomes larger in large
partonic system. In the present work, we did not observe any clear dependence of $V$ on rapidity because the trend of $V$ is almost constant. However we may consider more analysis for the detailed study of dependence of $V$ on rapidity in the future. Additionally, $V$ is observed to be mass dependent. Larger the mass of the particle,
smaller the $V$ which shows the early freeze-out of massive particles. This result is consistent with our previous results [4, 6, 13].
\\

{\section{Conclusions}}

We summarize here our main observations and conclusions.

(a) The transverse momentum spectra of positively and negatively charged
particles produced in inelastic proton-proton collisions in different rapidity
slices have been studied by the Blast Wave model with Boltzmann Gibbs statistics.
The results are well in agreement with the experimental data measured by the NA61/SHINE
Collaboration at CERN over an energy range from 20 GeV to 158 GeV.

(b) Kinetic freeze-out temperature increase with the increase of collision energy
due to large deposition of energy in the system at higher energies, and it decreases
with increase of rapidity because of less energy transfer in the system due to large penetration
between participant nucleons. It is also  observed that kinetic freeze-out temperature increases with mass.

(c) The transverse flow velocity is observed to have no dependence on rapidity and energy. However
it is dependent on the mass of particle. Massive particles have smaller $\beta_T$.

(d) There is no dependence of kinetic freeze-out volume observed on rapidity in the present work. However,
$V$ increase with increase of energy due to longer evolution time at higher energies. It is also
observed that volume differential scenario is observed as the massive particles have smaller $V$ and they freeze-out early.
\\

{\bf Acknowledgments}
This work is supported by the
National Natural Science Foundation of China (Grant
Nos. 11875052, 11575190, and 11135011).
We would also would like to acknowledge the support
of Ajman University Internal Research Grant NO. [DGSR Ref. 2020-IRG-HBS-01].
\\
\\
{\bf Author Contributions} All authors listed have made a
substantial, direct, and intellectual contribution to the work and
approved it for publication.
\\
\\
{\bf Data Availability Statement} This manuscript has no
associated data or the data will not be deposited. [Authors'
comment: The data used to support the findings of this study are
included within the article and are cited at relevant places
within the text as references.]
\\
\\
{\bf Compliance with Ethical Standards}
\\
\\
{\bf Ethical Approval} The authors declare that they are in
compliance with ethical standards regarding the content of this
paper.
\\
\\
{\bf Disclosure} The funding agencies have no role in the design
of the study; in the collection, analysis, or interpretation of
the data; in the writing of the manuscript, or in the decision to
publish the results.
\\
\\
{\bf Conflict of Interest} The authors declare that there are no
conflicts of interest regarding the publication of this paper.

\end{multicols}
\end{document}